\documentstyle[12pt,axodraw]{article}
\newcommand{\be}{\begin{equation}}
\newcommand{\ee}{\end{equation}}
\newcommand{\bea}{\begin{eqnarray}}
\newcommand{\eea}{\end{eqnarray}}
\newcommand{\al}{\alpha}

\newcommand{\gm}{\gamma}
\newcommand{\Gm}{\Gamma}

\newcommand{\ep}{\epsilon}

\newcommand{\dd}{\mbox{d}}

\newcommand{\nn}{\nonumber}
\newcommand{\uk}{\underline{k}}

\begin{document}
\parindent=1.5pc

\begin{titlepage}
\rightline{hep-ph/9907471}
\rightline{July 1999}
\bigskip
\begin{center}
{{\bf Problems of the Strategy of Regions } \\
\vglue 5pt
\vglue 1.0cm
{ \large V.A. Smirnov\footnote{E-mail: smirnov@theory.npi.msu.su} }\\
\baselineskip=14pt
\vspace{2mm}
{\em Nuclear Physics Institute of Moscow State University}\\
{\em Moscow 119899, Russia}
\vglue 0.8cm
{Abstract}}
\end{center}
\vglue 0.3cm
{\rightskip=3pc
 \leftskip=3pc
\noindent
Problems that arise in the application of general prescriptions of the
so-called strategy of regions for asymptotic expansions of Feynman
integrals in various limits of momenta and masses are discussed with
the help of characteristic examples of two-loop diagrams. The strategy
is also reformulated in the language of alpha parameters.
\vglue
0.8cm}

\end{titlepage}

\section{Introduction}

When analyzing the leading asymptotic behaviour in various limits a
standard strategy of regions is often used: instead of the integration
over the whole space of loop momenta only the integration over some
specific regions is considered. This strategy turns out to be possible
because the  information about the leading behaviour is somehow
encoded in this integration. A classical example of this procedure is
the analysis and summation of the leading logarithms in the Sudakov
limit \cite{Sud}. It was argued (and demonstrated for the threshold
expansion) in \cite{BS} that it is reasonable to use the strategy of
regions in the most general form and apply it to the whole asymptotic
expansion, i.e. for any powers and logarithms in an arbitrary limit.
In this generalized form, the strategy reduces to the following
prescriptions:

({\it i})
Consider various regions of the loop momenta and expand, in
every region, the integrand in Taylor series with respect to the
parameters that are considered small in the given region;

({\it ii})
Integrate the integrand expanded, in every region in its own way, over
the whole integration domain of the loop momenta;

({\it iii})
Put to zero any scaleless integral.

In the case of off-shell and off-threshold limits, this strategy leads
to the well-known explicit prescriptions \cite{offae,offae-proof} (see
a brief review \cite{aerep}) based on the strategy of subgraphs. Some
subtle points in the application of these general prescriptions of
strategy of regions to on-shell limits were discussed in
ref.~\cite{SR}. The purpose of this paper is to further discuss the
status of the strategy of regions using characteristic examples of
two-loop diagrams. We present two examples of vertex diagrams in
Section~2. In Section~3, we reformulate the strategy of regions in the
language of alpha parameters. In conclusion, we discuss the status of
the strategy of regions and present various recipes for its
application.

\section{Two examples}

Our first example is the diagram of Fig.~1 with the masses
$m_1=m_2=m_3=m_4=m, \; m_5=m_6=0$, the external momenta
$p_1^2=p_2^2=m^2$ and $(p_1-p_2)^2=-Q^2$ in the limit
$m/Q\to 0$.

\begin{center}
\begin{figure}[thb]
\begin{picture}(400,160)(-180,0)
\Line(10,100)(60,100)
\Line(10,150)(60,150)
\Line(10,100)(10,150)
\Line(60,150)(60,100)
\Line(60,100)(35,70)
\Line(35,70)(10,100)
\ArrowLine(35,50)(35,70)
\ArrowLine(10,150)(10,170)
\ArrowLine(60,170)(60,150)

\Text(35,45)[]{$p_1-p_2$}

\Text(10,175)[]{$p_1$}
\Text(60,175)[]{$p_2$}
\Text(  0,118)[]{3}
\Text( 25, 70)[]{1}
\Text( 73,118)[]{4}
\Text( 50, 70)[]{2}
\Text( 35,140)[]{5}
\Text( 35, 90)[]{6}

\Vertex(10,100){1.5}
\Vertex(60,100){1.5}
\Vertex(10,150){1.5}
\Vertex(60,150){1.5}
\Vertex(35,70){1.5}
\end{picture}
\vspace{-10pt}
\caption {
(a) Two-loop planar vertex diagram.}
\label{2l}
\end{figure}
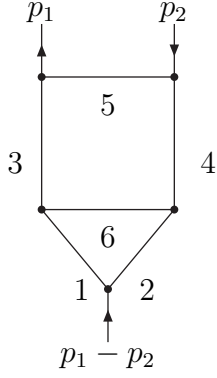
\end{center}

The Feynman integral can be written as
\be
 F_{1}(Q,m,\ep)
= \int  \int \frac{\dd^dk \dd^dl}{(l^2-2 p_1 l) (l^2-2 p_2 l)
(k^2-2 p_1 k) (k^2-2 p_2 k) k^2 (k-l)^2} \, .
\label{F1}
\ee
We use dimensional regularization \cite{dimreg} with $d=4-2\ep$.
When presenting our results we shall omit $i\pi^{d/2}$ per loop and,
when writing down separate contributions through  expansion in $\ep$,
we shall also omit $\exp(-\gm_E \ep)$ per loop ($\gm_E$ is the Euler
constant).

Let us choose, for convenience, the external momenta as follows:
\be
p_1 = \tilde{p}_1 +\frac{m^2}{Q^2} \tilde{p}_2 \, , \;
p_2 = \tilde{p}_2 +\frac{m^2}{Q^2} \tilde{p}_1 \, , \;
\tilde{p}_{1,2} = (Q/2,\mp Q/2,0,0)
\ee
so that  $p_i^2 = m^2 \, , \; \tilde{p}_i^2 = 0 \, , \;
2\tilde{p}_1 \tilde{p}_2= 2\tilde{p}_1 p_2 = Q^2$.
In the given limit, the following regions happen to be typical \cite{Co}:
\bea
\label{h}
\mbox{{\em hard} (h):} && k\sim Q\, ,
\\
\label{1c}
\mbox{{\em 1-collinear} (1c):} && k_+\sim Q,\,\,k_-\sim m^2/Q\, ,
\,\, \uk \sim m\,,
\\
\label{2c}
\mbox{{\em 2-collinear} (2c):} && k_-\sim Q,\,\,k_+\sim m^2/Q\, ,
\,\,\uk \sim m \, ,
\\
\label{soft}
\mbox{{\em soft} (s):} && k\sim m \, ,
\\
\label{us}
\mbox{{\em ultrasoft} (us):} && k\sim m^2/Q\, .
\eea
Here $k_{\pm} =k_0\pm k_1, \; \uk=(k_2,k_3)$. We mean by $k\sim Q$, etc.
that any component of $k_{\mu}$ is of order $Q$.

In the leading order, $1/Q^4$, we obtain the following
five contributions generated by the regions from this list:
(h-h), (1c-h), (2c-h), (1c-1c), (2c-2c). (We indicate regions
for the loop momenta $k$ and $l$ in (\ref{F1}) respectively
in the first and the second place.) In contrast to the diagram
with another distribution of the masses and assignments of the
external momenta, $m_1=\ldots=m_4=0,\, m_5=m_6=m\,, p_1^2 =p_2^2 =0$,
where the last four contributions are not regulated dimensionally
so that it is necessary to introduce an auxiliary analytic regularization
\cite{Sm2}, every term can be now considered separately,
and we symbolically have (2c-h)=(1c-h), (2c-2c)=(1c-1c).

The (h-h) region generates terms obtained by Taylor expanding the
integrand in the expansion parameter, $m$. In the leading order, this
is the value of the massless planar diagram at $p_1^2 =p_2^2 =0$
evaluated in \cite{Gons}:
\bea
C^{(1)}_{(h-h)} &=&  \frac{c^{(1)}_{(h-h)}}{(Q^2)^{2+2\ep}} \,, \nn \\
c^{(1)}_{(h-h)} &=&
\frac{1}{4 \ep^4} + \frac{5 \pi^2}{24 \ep^2}  +
  \frac{29 \zeta(3)}{6 \ep} + \frac{3 \pi^4}{32}
 +O(\ep) \, .
\label{h-h}
\eea

Then we have
\bea
C^{(1)}_{(1c-1c)}
&=& \int  \int \frac{\dd^dk \dd^dl}{(-2 \tilde{p}_1 l) (l^2-2 p_2 l)
(-2 \tilde{p}_1 k) (k^2-2 p_2 k) k^2 (k-l)^2} \, ,
\label{1c-1c}
\nn \\
C^{(1)}_{(1c-h)}
&=& \int  \int \frac{\dd^dk \dd^dl}{(l^2-2 \tilde{p}_1 l)
(l^2-2 \tilde{p}_2 l)
(-2 \tilde{p}_1 k) (k^2-2 p_2 k)}
\nn \\ &\times &
\frac{1}{k^2(l^2 - (2 \tilde{p}_1 k) (2 \tilde{p}_2 l)/Q^2 )} \, .
\label{1c-h}
\eea
Using the technique of alpha parameters and Mellin-Barnes representation
we obtain the following results:
\bea
C^{(1)}_{(1c-1c)} &=& \frac{c^{(1)}_{(1c-1c)}}{
(Q^2)^{2} (m^2)^{2\ep}} \, ,\nn\\
c^{(1)}_{(1c-1c)}&=&
-\frac{\pi^2}{24\ep^2} + \frac{5\zeta(3)}{4\ep} + \frac{\pi^4}{48}
+O(\ep)
 \, , \label{1c-1cRes} \\
C^{(1)}_{(1c-h)}
&=& \frac{c^{(1)}_{(1c-h)}}{(Q^2)^{2+\ep} (m^2)^{\ep}}\,,\nn\\
c^{(1)}_{(1c-h)} &=&
-\frac{1}{6\ep^4} - \frac{\pi^2}{6\ep^2}
- \frac{41\zeta(3)}{9\ep} - \frac{37\pi^4}{180}
 +O(\ep)
 \, .
\label{1c-hRes}
\eea

If we combine all these five contributions we shall obviously
obtain a wrong result because the poles of the fourth and the third
order will not cancel. It turns out that the following non-standard region
has to be included into the list:
\be
\mbox{{\em 1-ultracollinear} (1uc):}  \;\;\;\;\;
k_+\sim m^2/Q,\,\,k_-\sim m^4/Q^3\, ,
\,\, \uk \sim m^3/Q^2\,.
\label{2uc}
\ee
The missing contributions are (2uc-1c) and (1uc-2c) which are equal
to each other and easily evaluated by alpha parameters
for general $\ep$. In the leading order, we have
\bea
C^{(1)}_{(2uc-1c)}
&=& \int  \int \frac{\dd^dk \dd^dl}{(-2 \tilde{p}_1 l)
(l^2-2 p_2 l) (-2 p_1 k) (-2 \tilde{p}_2 k) k^2
(l^2 - (2 \tilde{p}_1 l) (2 \tilde{p}_2 k)/Q^2 )} \,  \nn\\
&=& \frac{c^{(1)}_{(2uc-1c)}}{(Q^2)^{2-2\ep} (m^2)^{4\ep}}\,, \nn\\
c^{(1)}_{(2uc-1c)} &=&
-\Gm(\ep)\Gm(2\ep)\Gm(3\ep) \Gm(-4\ep)
\,  \nn\\&=&
\frac{1}{24\ep^4} + \frac{5\pi^2}{48\ep^2}
+ \frac{7\zeta(3)}{18\ep}  + \frac{493\pi^4}{2880}
 +O(\ep) \, .
\label{2uc-1c}
\eea

Collecting all the seven contributions together
we observe that the poles of the third and the fourth order in $\ep$
cancel, and we come to the following result
\bea
(Q^2)^{2+2\ep} F_{1} (Q,m,\ep)
&\stackrel{\mbox{\small $Q \to \infty$}}{\mbox{\Large$\sim$}} &
\nn \\ && \hspace*{-30mm}
c^{(1)}_{(h-h)}
+ 2 c^{(1)}_{(1c-h)} \left(\frac{Q^2}{m^2}\right)^{\ep}
+ 2 c^{(1)}_{(1c-1c)} \left(\frac{Q^2}{m^2}\right)^{2\ep}
+ 2 c^{(1)}_{(2uc-1c)} \left(\frac{Q^2}{m^2}\right)^{4\ep}
\nn\\ && \hspace*{-30mm}
= \ln^2\frac{m^2}{Q^2} \frac{1}{2\ep^2}
-   \left(\frac{5}{6} \ln^3\frac{m^2}{Q^2}
+ \frac{\pi^2}{3} \ln\frac{m^2}{Q^2}
+ \zeta(3)\right) \frac{1}{\ep}
\nn \\ && \hspace*{-30mm}
+  \frac{7}{8}\ln^4\frac{m^2}{Q^2}
+ \frac{4\pi^2}{3} \ln^2\frac{m^2}{Q^2}
+ \zeta(3)\ln\frac{m^2}{Q^2}  + \frac{\pi^4}{15}
+O(\ep) \, .
\eea
with the proper coefficient at the double pole which can be evaluated
starting from the full diagram.

Note that in the case of the non-planar diagram in the considered
limit, with $p_1^2=p_2^2=m^2$, there are exactly the same problems
with dimensional regularization as in the limit with $p_1^2=p_2^2=0$
\cite{Sm2,SR}.

The second example is the same diagram Fig.~1 with the masses
$m_1=m_3=M,\; m_2=m_4=m, \; m_5=m_6=0$, the external momenta
$p_1^2=M^2, \; p_2^2=m^2$ and $Q^2=(p_1-p_2)^2=0$ in the limit
$m/M\to 0$.
The corresponding Feynman integral $F_{2} (M,m,\ep)$
has literally the same form as (\ref{F1}), with
the other assignments for the external momenta.
Let us choose the external momenta as $p_1=(M,\vec{0}),
p_2= M n_1 + \frac{m^2}{M} n_2$,
with $n_{1,2} = (1/2,\mp 1/2,0,0)$.
We have $2n_{1,2} k=k_{\pm}$ for any $d$-vector $k$.

No `unusual' regions are relevant here. In the leading order, there
are four non-zero contributions, corresponding to (h-h),
(2c-h),(2c-2c) and (us-2c) regions. (Here the same characterization of
the regions (3)--(7)
in terms of the components $k_{\pm}$ and $\uk$, with the
substitution $Q\to M$, is implied.) The (h-h) contribution is obtained
by expanding the integrand in Taylor series in $m^2$. Using again the
technique of alpha parameters and Mellin-Barnes representation we come
to the following result:
\bea
C^{(2)}_{(h-h)} &=&
\frac{c^{(2)}_{(h-h)}}{(M^2)^{2+2\ep}}\, ,\nn\\
c^{(2)}_{(h-h)} &=&
\frac{1}{12 \ep^4} +\frac{\pi^2}{12 \ep^2}   + \frac{91 \zeta(3)}{36
\ep} + \frac{179 \pi^4}{1440} +O(\ep) \, . \label{h-h-(2)}
\eea

The contribution of the (2c-2c) region is obtained due
to the following prescriptions:

(a) Expand the propagators $1/(k^2-2p_1k)$ and $1/(l^2-2p_1l)$
in Taylor series respectively in $k^2$ and $l^2$,

(b) Expand each resulting term, which is a function of three
kinematical invariants, $p_1^2=M^2, p_2^2=m^2$ and $2p_1p_2=m^2+M^2$,
in a Taylor series at the point $p_1^2=0$ and $2p_1p_2=M^2$
(do not touch $p_2^2=m^2$).

It might seem that we expand in the large mass $M$ but this is
just an illusion. The (2c-h) contribution is of an intermediate
character.
In the leading order, we have
\bea
\hspace*{-7mm} C^{(2)}_{(2c-2c)}
&=&
\int  \int \frac{\dd^dk \dd^dl}{(-2 P_2 l) (l^2-2 p_2 l)
(-2 P_2 k) (k^2-2 p_2 k) k^2 (k-l)^2} \, ,
\label{2c-2c-(2)}
\nn \\
\hspace*{-7mm} C^{(2)}_{(2c-h)}
&=&
\int  \int \frac{\dd^dk \dd^dl}{(l^2-2 p_1 l) (l^2-2 P_1 l)
(-2 P_2 k) (k^2-2 p_2 k) k^2
(l^2-(2n_1 l)\frac{2 P_2 k}{M}  ) } \, ,
\label{2c-h-(2)}
\eea
where $P_1=M n_1,\;P_2=M n_2$.
These contributions are evaluated by the same techniques as before,
with the following results:
\bea
C^{(2)}_{(2c-2c)} &=&
\frac{c^{(2)}_{(2c-2c)}}{(M^2)^{2} (m^2)^{2\ep}} \,, \nn \\
c^{(2)}_{(2c-2c)}&=& c^{(1)}_{(2c-2c)}
=c^{(1)}_{(1c-1c)}
 \, , \label{2c-2cRes-(2)} \\
C^{(2)}_{(2c-h)} &=&
\frac{c^{(2)}_{(2c-h)}}{(M^2)^{2+\ep} (m^2)^{\ep}} \, , \nn\\
c^{(2)}_{(2c-h)}  &=&
-\frac{1}{8 \ep^4} - \frac{7 \pi^2}{48 \ep^2}
 -  \frac{31 \zeta(3)}{6 \ep} - \frac{871 \pi^4}{2880} + O(\ep)
 \, .
\label{2c-hRes-(2)}
\eea

The (us-2c) contribution
\be
C^{(2)}_{(us-2c)} =
\int  \int \frac{\dd^dk \dd^dl}{(-2 P_2 l) (l^2-2 p_2 l)
(-2 p_1 k) (-2 P_1 k) k^2
(l^2-(2n_1 k)\frac{2 P_2 l}{M}  ) } \,
\ee
is evaluated in gamma functions, with
a result which is closely related to the (2uc-1c) and (1uc-2c)
contributions in Example~1:
\bea
C^{(2)}_{(us-2c)}
&=& \frac{c^{(2)}_{(us-2c)}}{(Q^2)^{2-\ep} (m^2)^{3\ep}}\,, \nn\\
c^{(2)}_{(us-2c)} &=& c^{(1)}_{(2uc-1c)}
= c^{(1)}_{(1uc-2c)}  \, .
\label{us-2c}
\eea

Collecting all the four contributions together
we observe that the poles of the third and the fourth order in $\ep$
cancel, and we come to the following result
\bea
(M^2)^{2+2\ep} F_{2} (M,m,\ep)
&\stackrel{\mbox{\small $M \to \infty$}}{\mbox{\Large$\sim$}} &
\nn \\ && \hspace*{-30mm}
c^{(2)}_{(h-h)}
+ c^{(2)}_{(2c-h)} \left(\frac{M^2}{m^2}\right)^{\ep}
+ c^{(2)}_{(2c-2c)} \left(\frac{M^2}{m^2}\right)^{2\ep}
+ c^{(2)}_{(us-2c)} \left(\frac{M^2}{m^2}\right)^{3\ep}
\nn\\ && \hspace*{-30mm}
= \ln^2\frac{m^2}{M^2} \frac{1}{8\ep^2}
-   \left(\frac{1}{6} \ln^3\frac{m^2}{M^2} + \frac{\pi^2}{12} \ln\frac{m^2}{M^2}
+ \zeta(3)\right) \frac{1}{\ep}
\nn \\ && \hspace*{-30mm}
+  \frac{13}{96}\ln^4\frac{m^2}{M^2}  + \frac{5\pi^2}{16} \ln^2\frac{m^2}{M^2}
+ \frac{3\zeta(3)}{2}\ln\frac{m^2}{M^2}  + \frac{\pi^4}{72}
+O(\ep) \, .
\eea
with the proper coefficient at the double pole which can be evaluated
starting from the full diagram.

\section{The language of the alpha representation}

There are also contributions of the (h-2c) region (in both examples)
and (h-1c) region (in the first example) which can be evaluated
in gamma functions for general $\ep$ and start from the next-to-leading
order, $m^2$. Consider now another choice of the loop momenta where
the momentum $l$ is chosen as the momentum of the 2nd line. Then
we can recognize a non-zero contribution from the (h-s) region ($k$
is hard and this new momentum $l$ is soft). However this is nothing
but a double counting because these contributions are identical:
this can be seen by analyzing this shift of the variables $l\to l-p_2$.
This example shows that one has to be careful when testing different
choices of the loop momenta. Of course, there is no sense of choosing
different assignments of the loop momenta for the hard regions. But
for other types of regions certain choices of the loop momenta
are crucial --- see, e.g., two-loop examples in the case of
the threshold expansion in \cite{BS}, in particular, in the
case of the ultrasoft regions. Here is an another example where
a different assignment of the loop momenta is important: consider
Fig.~1 with $m_1=\ldots=m_4=0,\, m_5=m_6=m\,, p_1^2 =p_2^2 =0$ and
$m^2/(p_1-p_2)^2\to 0$. Then it is necessary to take into account the (h-s)
region with hard $k$ and the soft momentum of the 6th line \cite{Sm2}.
(It also contributes from the next-to-leading order.)

There is a possibility to completely avoid such double counting by
turning to the alpha representation and using a similar strategy of
regions in the language of alpha parameters. This representation for a
$h$-loop scalar diagram $\Gm$ with powers of propagators $a_l$ has the
form
\bea
F_{\Gm}(q_1,\ldots,q_n) & =&
(-1)^a\frac{ i^{a+h (1-d/2)}\pi^{h d/2}}{\prod_l \Gm(a_l)}
\nn \\ && \hspace*{-20mm}
\times \int_0^\infty \dd\al_1 \ldots\int_0^\infty\dd\al_L
\prod_l \al_l^{a_l-1}
D^{-d/2} e^{i A/D-i\sum m_l^2 \al_l} \; ,
\label{alpha-d}
\eea
where
\bea
D& = & \sum_T \prod_{l\bar{\in} T} \al_l
 \;,
\label{Dform}  \\
A&=& \sum_T \left[\prod_{l\bar{\in} T} \al_l  \left(\sum_i q_i\right)^2\right]
 \; .
\label{Aform}
\eea
In (\ref{Dform}), the sum runs over trees of the given graph, i.e.
maximal connected subgraphs without loops, and, in (\ref{Aform}),
over 2-trees, i.e. subgraphs that do not involve loops
and consist of two connectivity components. The sum of momenta present
in (\ref{Aform}) goes over the external momenta that flow
into one of the connectivity components of the 2-tree $T$.
The products of the alpha parameters involved are taken
over the lines that do not belong to the given tree $T$.
In the non-scalar case, there appear additional factors in the integrand
of the alpha representation --- see, e.g., \cite{BM}.

The strategy of regions in the alpha representation is formulated
in the same way as in the integrals in the loop momenta. One has to
consider alpha parameters to be of different order measured in terms
of given masses and kinematical invariants.
Let us illustrate the new language
using the two above examples. The function (\ref{Dform}) is the same
in both cases
\be
D = (\al_1+\al_2) (\al_3+\al_4+\al_5)
+\al_6 (\al_1+\al_2+\al_3+\al_4+\al_5) \, .
\ee
The functions (\ref{Aform}) are
\bea
A_1 & = &
[ (\al_1+\al_3) (\al_2+\al_4)   \al_6 + (\al_1+\al_2) \al_3\al_4
+ \al_1\al_2 (\al_3+\al_4+\al_5)] Q^2
\nn \\ &+ &
[(\al_1+\al_2) ( (\al_3+\al_4) (\al_1+\al_2+\al_3+\al_4)
\nn \\ & +&
(\al_1+\al_2)\al_5) + (\al_1+\al_2+\al_3+\al_4)^2\al_6 ] m^2 \, ,
\nn \\
A_2 & = &
[M^2 (\al_1+\al_3) + m^2 (\al_2+\al_4)]
[(\al_1+\al_2+\al_3+\al_4)\al_6+(\al_1+\al_2)(\al_3+\al_4) ]
\nn \\ & +&
[M^2 \al_1 + m^2 \al_2] (\al_1+\al_2) \al_5
  \, .
\eea

The contribution of the momentum space (h-h) region is obtained when
considering all the $\al_l$ as being of the same order. (Note that the
functions in the alpha representation are homogeneous so that it
suffices to take into account only the relative order of the alpha
parameters.) Thus the form $D$ is not expanded
and the part of the exponent proportional to $m^2$ is expanded
in Taylor series in $m$.

In Example~1, we reproduce the contributions of the previously
considered momentum space regions as follows:
\bea
\mbox{(1c-h)} & \to & \{\al_{4,5} \sim m^0\,, \; \al_{1,2,3,6} \sim m^2 \} ;
\nn \\
\mbox{(1c-1c)} & \to & \{\al_{2,4,5,6} \sim m^0\,, \; \al_{1,3} \sim m^2 \} ;
\nn \\
\mbox{(h-1c)} & \to & \{\al_{2} \sim m^0\,, \; \al_{1,3,4,5,6} \sim m^2 \} ;
\nn \\
\mbox{(2uc-1c)} & \to &
\{\al_{5} \sim m^0\,, \; \al_{3} \sim m^2 \,, \;
\al_{2,4,6} \sim m^4\,, \; \al_{1} \sim m^6 \} .
\eea
The alpha parameters have dimension $m^{-2}$ but to simplify
relations we put $Q$ and $M$ to one and express
all the magnitudes in powers of $m$.
The rest of the regions is obtained by permutations.

In Example~2, we have
\bea
\mbox{(2c-h)} & \to & \{\al_{4,5} \sim m^0\,, \; \al_{1,2,3,6} \sim m^2 \} ;
\nn \\
\mbox{(2c-2c)} & \to & \{\al_{2,4,5,6} \sim m^0\,, \; \al_{1,3} \sim m^2 \} ;
\nn \\
\mbox{(h-1c)} & \to & \{\al_{2} \sim m^0\,, \; \al_{1,3,4,5,6} \sim m^2 \} ;
\nn \\
\mbox{(us-2c)} & \to &
\{\al_{5} \sim m^0\,, \; \al_{2,3,4,6} \sim m^2 \,, \;
\al_{1} \sim m^4 \} .
\eea

Note that here we do not have risk to perform double counting
of the (h-2c) and (h-s) (with another choice of the second loop momentum)
regions which was observed above. Moreover, the description of the regions
in the alpha parametric language is manifestly Lorentz-invariant.
However, a visible drawback of this language is rather non-trivial
description of the (us-2c) and especially (2uc-1c) regions. To find them
one would need to consider an enormous quantity of various
possibilities.

Let us turn to one more example: the on-shell double box diagram shown in
Fig.~2. We have $p_i^2=0,\; i=1,2,3,4$. It was evaluated analytically in
ref.~\cite{DB}. An explicit analytical algorithm for on-shell double
boxes with arbitrary numerators and integer powers of propagators was
presented in ref.~\cite{SV}.
\begin{figure}[hbt]
\centering
\begin{picture}(160,60)(0,2)
\put(20,10){\line(1,0){120}}
\put(20,50){\line(1,0){120}}
\put(40,10){\line(0,1){40}}
\put(80,10){\line(0,1){40}}
\put(120,10){\line(0,1){40}}
\put(40,10){\circle*{3}}
\put(40,50){\circle*{3}}
\put(120,10){\circle*{3}}
\put(120,50){\circle*{3}}
\put(80,10){\circle*{3}}
\put(80,50){\circle*{3}}
\put(6,9){$p_1$}
\put(6,49){$p_2$}
\put(146,9){$p_3$}
\put(146,49){$p_4$}
\put(32,27){\small $5$}
\put(72,27){\small $6$}
\put(112,27){\small $7$}
\put(59,0){\small $3$}
\put(99,0){\small $1$}
\put(59,53){\small $4$}
\put(99,53){\small $2$}
\end{picture}
\caption{Double box diagram.}
\end{figure}
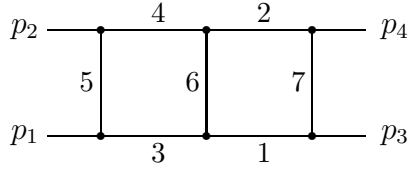
In \cite{SV} these results were also checked against first terms of
the expansion in $t/s$, where $s=(p_1+p_2)^2$ and $t=(p_1+p_3)^2$
are Mandelstam variables, with the help of the strategy of regions.
The external momenta were chosen as
\be
p_{1,2} = (\mp Q/2,Q/2,0,0), \;
r\equiv p_1+p_3 = (-t/Q,0,\sqrt{-t+t^2/Q^2},0) \, ,
\ee
where $s=-Q^2$. The non-zero contributions to the asymptotic
expansion in the limit $t/s\to 0$ are (h-h), (1c-1c) and (2c-2c).
(The description of the collinear regions is given by
eqs.~(\ref{1c}) and (\ref{2c}),
with the replacement $m^2\to -t$.)
The (1c-1c) and (2c-2c) contributions are not individually regularized
by dimensional regularization; a natural way to deal with regularized
quantities is to introduce and auxiliary analytic regularization
which is eventually switched off in the sum.

When analyzing the limit $s/t\to 0$, it is reasonable to rotate the
diagram and perform the substitutions $t\leftrightarrow s\, ,
p_1\to p_2\,,\; p_2\to p_4$, etc.
The non-zero contributions to the asymptotic
expansion in the limit $s/t\to 0$ are (h-h), (1c-1c), (2c-2c) and (2c-1c).
Here the introduction of an auxiliary analytic regularization is also
reasonable. However the corresponding poles in the analytic regularization
parameter happen to be here up to the second order, as in an example
of a non-planar vertex diagram in ref.~\cite{SR}.

Let us now describe the contributions to the asymptotic
expansion in the limits $t/s \to 0$ and $s/t \to 0$ in the
language of alpha parameters.
The functions (\ref{Dform}) and (\ref{Aform}) for the double box
are
\bea
D_3&=&(\al_1+\al_2+\al_7) (\al_3+\al_4+\al_5)
+\al_6 (\al_1+\al_2+\al_3+\al_4+\al_5+\al_7) \;,
\label{DformDB}
\\
A_3 &=& [\al_1\al_2 (\al_3+\al_4+\al_5) + \al_3\al_4(\al_1+\al_2+\al_7)
+\al_6 (\al_1+\al_3)( \al_2+\al_4) ] s
\nn \\ &&
+ \al_5\al_6\al_7 t
\label{AformDB} \; .
\eea
The (h-h) contributions are reproduced by considering
all the alpha parameters of the same order. Let us again turn to
dimensionless alpha parameters by putting $s=-1$ ($t=-1$)
in the first (second) limit.
In the limit $t/s\to 0$, we reproduce the contributions of the previously
considered collinear momentum space regions as follows:
\bea
\mbox{(1c-1c)} & \to & \{\al_{2,4,5,6,7} \sim t^0\,, \;
\al_{1,3} \sim -t \} ;
\nn \\
\mbox{(2c-2c)} & \to & \{\al_{1,3,5,6,7} \sim t^0\,, \;
\al_{2,4} \sim -t \} \, ,
\eea
and, in the limit $t/s\to 0$, we have
\bea
\mbox{(1c-1c)} & \to & \{\al_{1,2,3,4,6,7} \sim s^0\,, \;
\al_{5} \sim -s \} ;
\nn \\
\mbox{(2c-2c)} & \to & \{\al_{1,2,3,4,5,6} \sim s^0\,, \;
\al_{7} \sim -s \} ;
\nn \\
\mbox{(2c-1c)} & \to & \{\al_{1,2,3,4,5,7} \sim s^0\,, \;
\al_{6} \sim -s \} \,.
\eea
We see that, in this example, the description of the collinear
contributions in the language of alpha parameters is certainly
simpler than in the momentum space.

\section{Discussion}

Let us realize that the very word `region' is understood
in the `physical' sense. In fact, it indicates relations between
components of the loop momenta expressed in terms of the given
masses and kinematical invariants. This is clearly not the mathematical
sense where the region is determined by inequalities.
We even do not bother about the decomposition of unity, i.e.
that our initial integral in the whole space of the loop momenta
is decomposed into a sum of integrals over all the possible regions
which, presumably, have zero measure in the intersection of any pair,
with their union being the whole integration space.

But the most non-trivial step in the strategy of regions is, probably,
the last one when all the scaleless integrals are  put to zero. Note
that this step generally does not refer to the use of dimensional
regularization. For example, some integrals generated by potential
contributions within threshold expansion \cite{BS} were not at all
regularized. Still the rule to put such scaleless integrals to zero
was experimentally checked through examples.

Up to now there are no counterexamples that would show that the
strategy of regions generally does not work. Still mathematical proofs
in the general case are also absent. An indirect confirmation of the
strategy is the fact that such proof is indeed available
\cite{offae-proof} in the case of off-shell and off-threshold limits.
(These are limits typical for Euclidean space.
For this class of regions, it suffices to consider any loop momentum
to be either hard or soft.) The decisive point of this proof is the
analysis of convergence of the Feynman integrals. To be more precise,
this is the resolution of singularities of integrands which is usually
performed in the alpha representation. To resolve singularities of
(\ref{alpha-d}) the whole integration domain is decomposed into
sectors. These are either $\al_1\leq\ldots\leq\al_L$ plus sectors
obtained by permutations, or more advanced sectors associated with
one-particle-irreducible subgraphs (and their infrared analogs
\cite{Rstar}) of the given graph. After a suitable change of variables
in each sector, the integrand is factorized and the analysis of
convergence reduces to power counting in one-dimensional integrals.

To carry out the analysis of convergence for a given on-shell or
threshold limit we first need to invent appropriate sectors
and sector variables which would provide the factorization of
the corresponding integrand. At least the sectors mentioned above
are here insufficient --- this can be seen in one-loop examples.
It should be also stressed that, for each on-shell or threshold limit,
this problem of resolution of singularities should be solved
separately, with sectors and sector variables that
are specific for it.

Suppose that we have to expand a Feynman
diagram in some limit typical for Minkowski space. The first step is
to find all relevant regions that generate non-zero contributions. (For
the off-shell limits, there is no problem here:
we use the well-known graph-theoretical language for writing down
prescription for expanding the given Feynman diagram and just list all
relevant subgraphs that are taken from a certain family. This task can
be even done by computer. In some partial cases, the
prescriptions for on-shell limits can be also formulated in a
graph-theoretical language \cite{on-shell}.) The problem to
successfully go through this step is a matter of experience and (both
physical and mathematical) intuition. The physical flavour of the
problem is a correspondence of the class of the regions to certain
operators and subsequent translating the prescriptions for the
asymptotic expansion into the operator language.

When testing various regions it is necessary to be aware of possible
double counting. The combination of this search of the relevant
regions both in momentum space and alpha integrals looks rather
reasonable. But how can we decide that we have found all the
contributions to the asymptotic expansion? Unfortunately, there is no
definite answer to this question. At least we can check our results
by comparing them with one-loop examples where explicit analytical
results can be obtained. Sometimes comparisons with analytical
two-loop results are also available --- see, e.g., examples in
\cite{DB,SR}.
Another important partial check is the cancellation of poles in $\ep$,
up to a certain order, and the analysis of the coefficient at the
highest pole which can be evaluated by an independent method.

Thus, if the situation with the poles is unsatisfactory there are at least
two options:

(a) to decide that the strategy of regions breaks
down in the given limit;

(b) to look for missing regions.

Since the first option has been never realized, it looks reasonable
to stay always optimistic and continue the search of regions.
The regions can be rather non-trivial indeed:  the ultracollinear region
which can be symbolically described as $1uc=(m^2/Q^2)\times 1c$
is a strange example. By the way, any
{\em 1-ultra-\ldots-ultracollinear} region
$1uc=(m^2/Q^2)^h\times 1c$ is also a reality for an arbitrary $h$.
Consider the $h$-loop ladder diagram in the limit of Example~1.
Then the contributions of
(2u$^{h-1}$c-1u$^{h-2}$c-\ldots-1c) and
(1u$^{h-1}$c-2u$^{h-2}$c-\ldots-2c)
regions are non-zero (here $h$ is supposed to be even, for
definiteness).

But imagine now that our terms of expansion satisfy the check of poles.
The success is not yet guaranteed because we cannot exclude
the existence of a region that enters with simple poles in $\ep$
or even without poles which is insensitive to this check.
Then one could use numerical checks with numerical evaluation of the
initial diagram (and, of course, stay optimistic).

After this advice, let us stress that there is at least
an example of the on-shell double box diagrams when it turns out
easier to evaluate them analytically, rather than expand
them up to a desired order \cite{DB,SV}. So, the last advice is
to try to evaluate the diagrams analytically, without any expansion.
Still in this case, we can use the strategy of regions
for crucial checks --- see, e.g., \cite{SV}.

\vspace{2mm}

{\em Acknowledgments.}
I am grateful to M.~Beneke and A.~Hoang for useful discussions.
The work was supported by the Volkswagen Foundation, contract
No.~I/73611, and by the Russian Foundation for Basic Research,
project 98--02--16981.


\begin{thebibliography}{99}

\bibitem{Sud}
V.V. Sudakov, {\em Zh. Eksp. Teor. Fiz.} 30 (1956) 87.

\bibitem{BS}
M. Beneke and V.A.~Smirnov, {\em Nucl. Phys.} B522 (1998) 321.

\bibitem{offae}
S.G.~Gorishny, preprints JINR E2--86--176, E2--86--177 (Dubna 1986);
{\em Nucl. Phys.} B319 (1989) 633;
K.G.~Chetyrkin, {\em Teor. Mat. Fiz.} 75 (1988) 26; 76 (1988) 207;
K.G.~Chetyrkin, preprint MPI-PAE/PTh 13/91 (Munich, 1991).

\bibitem{offae-proof}
V.A.~Smirnov, {\em Commun. Math. Phys.} 134 (1990) 109;
V.A.~Smirnov, {\em Renormalization and asymptotic expansions}
(Birkh\"{a}user, Basel, 1991).

\bibitem{aerep}
V.A.~Smirnov, {\em Mod. Phys. Lett.} A 10 (1995) 1485.

\bibitem{SR}
V.A.~Smirnov and E.R. Rakhmetov, {\em Teor. Mat. Fiz.} 120 (1999) 64.

\bibitem{dimreg}
G.~'t Hooft and M.~Veltman, {\em Nucl.~Phys.} B44 (1972) 189;
C.G.~Bollini and J.J.~Giambiagi, {\em Nuovo Cim.} 12B (1972) 20.

\bibitem{Co}
J.C. Collins, in {\em Perturbative QCD}, ed. A.H. Mueller, 1989, p.~573.

\bibitem{Sm2}
V.A.~Smirnov, {\em Phys. Lett.} B404 (1997) 101;
Proceedings of
5th International Conference on Physics Beyond the Standard Model
(Balholm, Norway, 29 April - 4 May 1997), p.~354. AIP, 1997.
(hep-ph/9708423).

\bibitem{Gons}
R.J.~Gonsalves, {\em Phys.~Rev.} D28 (1983) 1542.

\bibitem{BM}
P. Breitenlohner and D. Maison, {\em Commun. Math. Phys.} 52 (1977) 39.

\bibitem{DB}
V.A. Smirnov, hep-ph/9905323, to appear in {\em Phys. Lett. B}.

\bibitem{SV}
V.A. Smirnov and O.L. Veretin, hep-ph/9907385.

\bibitem{Rstar} K.G.~Chetyrkin and V.A.~Smirnov, Phys. Lett.
{\bf 144B} (1984) 419.

\bibitem{on-shell}
V.A.~Smirnov, {\em Phys. Lett.} B394 (1997) 205;
A.~Czarnecki and V.A.~Smirnov, {\em Phys. Lett.} B394 (1997) 211.

\end{thebibliography}
\end{document}